\documentclass{article}

\usepackage{amsmath,amssymb,bbm}

\usepackage{epic}
\usepackage{amssymb,amsfonts,amsmath}

\usepackage{color}
\usepackage{array}
\usepackage{tabularx}
\usepackage{float}
\usepackage[pdftex]{graphicx}
\usepackage[noblocks]{authblk}

\begin{document}

\title{Phase diagram of a Schelling segregation model}

\author{L. Gauvin \footnote{Corresponding author, laetitia.gauvin@lps.ens.fr}}
\author{J. Vannimenus}
\affil{Laboratoire de Physique Statistique (UMR
  8550 CNRS-ENS-Paris 6-Paris 7), Ecole Normale Sup\'erieure, Paris,
  France}
\author{J.-P. Nadal$^{1,}$}
\affil{Centre d'Analyse et de Math\'ematique Sociales (UMR 8557
CNRS-EHESS), Ecole des Hautes Etudes en Sciences Sociales, Paris, France}
\date{26 March 2009}

\maketitle

\begin{abstract}  The collective behavior in a variant of Schelling's segregation model is characterized with methods borrowed from statistical physics, in a context where their relevance was not conspicuous. A measure of segregation based on cluster geometry is defined  and several quantities analogous to those used to describe physical lattice models at equilibrium are introduced. This physical approach allows to distinguish quantitatively several regimes and to characterize the transitions  between them, leading to the building of a phase diagram. Some of the transitions evoke empirical sudden ethnic turnovers. We also establish links with 'spin-1' models in physics. Our approach provides generic tools to analyze the dynamics of other socio-economic systems.\\


\end{abstract}

\section*{Introduction}
In the course of his study of the segregation effects observed in many social situations, Thomas Schelling introduced in the $1970$'s \cite{schelling1971,schelling1978} a model that has attracted a lot of attention ever since, to the point that it may now be considered an archetype in the social sciences. The success of the Schelling model is due to several factors: It was one of the first models of a complex system to show emergent behavior due to interactions among agents; it is very simple to describe, yet its main outcome - that strong segregation effects can arise from rather weak individual preferences - came as a surprise and proved robust with respect to various more realistic refinements; as a consequence, it has possibly far-reaching implications for social and economic policies aiming at fighting urban segregation, considered a major issue in many countries (for recent discussions of the social relevance of the model, see \cite{PancsVriend} and \cite{Clark}).\

An important recent development is the realization that there exists a striking kinship between this model and various physical models used to describe surface tension phenomena \cite{Kirman} or phase transitions and clustering effects, such as the Ising model \cite{StS,Marsili,Odor}. This connection is not a rigorous correspondence, still it is more than a mere analogy. It gives novel insight into the behavior of the Schelling model, and it suggests more generally that socio-economic models may  be fruitfully attacked drawing from the toolbox of statistical physics \cite{Marsili,CastFL}. Indeed, physicists developed during the last decades powerful methods for situations where
 obtaining analytical results seems out of reach.
These rely on the quantitative analysis of computer simulation results, guided by some general principles. They are well suited to complex systems such as those encountered in the social sciences and should in particular prove powerful in conjunction with agent-based modeling\footnote{The 'hand-made simulations' done by T. Schelling by moving pawns on a chessboard can be considered as the first agent-based simulations ever done in social science.},
a growingly popular approach \cite{Sackler}.\

A key physicist strategy is to characterize the system under study by the {\it phase diagram} which gives, in the space of control parameters, the boundaries separating domains of different qualitative behaviors. Each type of behavior is qualified by the order of magnitude of a small set of macroscopic quantities, the so-called ``order parameters''. The main difficulties are to correctly identify the relevant set of order parameters and the associated qualitative behaviors of the system, and to locate and characterize the boundaries, on which ``phase transitions'' occur in a smooth or discontinuous way.
In the present paper, we show how several of the methods evoked above can be adapted and applied to
the building of the phase diagram of social dynamics models, taking as paradigm the Schelling segregation model.
More precisely, we illustrate our approach on a particular variant of the Schelling model, 
where the basic variables are the tolerance (to be defined precisely below) and the density of vacant sites.
We introduce two order parameters which provide a relevant measure of global segregation, a surrogate of the energy,
and analogues of the susceptibility and the specific heat. We also introduce a real-space renormalization method suited to situations with a high density of vacancies. Our analysis shows the existence in the phase diagram of the model of sharp transitions, where relevant quantities have singularities, and we discuss the nature of these transitions. They separate different types of behavior - segregated, mixed, or "frozen" -,  in agreement with qualitative observations based on pictures of simulated systems \cite{Kirman}. In addition we make contact with the Blume-Emery-Griffiths \cite{BEG} and Blume-Capel \cite{Blume,Capel} models, which are spin-1 models much studied in relation with binary mixtures containing mobile vacancies and which show a richer behavior than the simple Ising model, with both discontinuous and continuous transitions.\
One should insist that most variants of the Schelling segregation model are of kinetic nature: their dynamics cannot be described as the relaxation to an equilibrium characterized by some energy function, except for specific variants (for such exceptions, see e.g. \cite{SiVaWe,inprep} and the discussion below on the links with spin models). Nevertherless, the tools and quantities we introduce by analogy with equilibrium statistical mechanics appear to be quite efficient to characterize the model behaviors. They are sufficiently general and relatively simple to be adaptable to a large variety of social and economic models,
as long as these involve interacting agents living in a discrete space
or more generally on a social network.

\section{Model and qualitative analysis} 

In Schelling's original model \cite{schelling1971} agents of two possible colors  
are located on the sites of a chessboard. Each color corresponds to members of one of  two  homogeneous groups which differ for example by their race, their wealth, etc... A fraction of the sites are blanks, the agents of both colors may move to these vacancies. The neighborhood of an agent comprises the eight nearest and next-nearest sites (Moore neighborhood). If less than $\frac{1}{3}$  of an agent's neighbors belong to his group, he is discontent -- in economic terms his utility is $0$; otherwise he is satisfied - his utility is  $1$. 
Starting from  random initial configurations  Schelling displaced discontent agents onto the closest satisfactory vacant sites,  
if possible. He observed that the system always reached a segregated state, where large clusters of same-color agents were formed. The crucial point is that segregation appears as an emergent phenomenon, in the sense that the collective effect is much stronger than what would be naively expected, as individual agents are happy to live in a mixed neighborhood. This phenomenon proves robust: a similar outcome, with some caveats,  is found in variants of the model, even when the utility function is non-monotonous with the fraction of similar neighbors \cite{PancsVriend,Jensen}.\\
The model we consider is a variant of the original Schelling model: The agents are satisfied with their neighborhood if it is constituted of a number of  \emph{unlike} agents $N_d$ lower than (or equal to) a fixed proportion $T$ of all the agents in the neighborhood. The parameter $T$ is called the tolerance \cite{schelling1971}. Since a higher tolerance allows for a larger number of configurations of satisfied agents (higher entropy), this parameter may be thought of as a temperature-like variable. 
We will be guided by this qualitative correspondence -- although in a different way than in  \cite{StS}, where a direct analogy is made with the Ising model. The other control parameter of the model is the vacancy density $\rho$. The randomly chosen agents move one by one to \emph{any} vacancy which has a satisfying neighborhood
- this is equivalent to long-range diffusion in physical terms. If no vacancy fits for some agents the latter respectively move back to their initial position. This dynamics is repeated until configurations are reached where the number of satisfied agents is almost stable. Note that in the present variant satisfied agents can also move, not just discontent ones. That rule introduces some noise in the dynamics and is useful to avoid a particularity of the original Schelling model noted in \cite{StS,SiVaWe}, namely that the system may end up in states where the clusters are large but finite, so that strictly speaking no large-scale segregation occurs. We will see later how the intensity of this noise is actually correlated with the tolerance level $T$. Note also that the global utility may decrease at times during the process, as the gain for the moved agent can be less than the net loss for his old plus new neighbors \cite{schelling1978}. \
Let us emphasize that only a finite number of values of $T$ are meaningful, namely $\frac{1}{8}, \frac{1}{7}..., \frac{6}{7}, \frac{7}{8}$. They correspond to the maximal  number of tolerated different neighbors divided by the actual number of occupied sites in the neighborhood. Any other value for $T$ is thus equivalent to the closest inferior meaningful value.

\subsection{ Numerical Simulations.}
Our simulations were performed on a $L \times L$ lattice ($L=50$, unless otherwise specified) with free boundary conditions. An initial configuration was randomly generated such that the vacancies and the two types of agents were fully mixed. 
Then the evolution followed the rules described above. In the simulations one time step corresponds to one attempted move per occupied site on average, the usual definition of a Monte-Carlo step.

\begin{figure}[h]
 \begin{center} 
 \includegraphics[width=6.5cm] {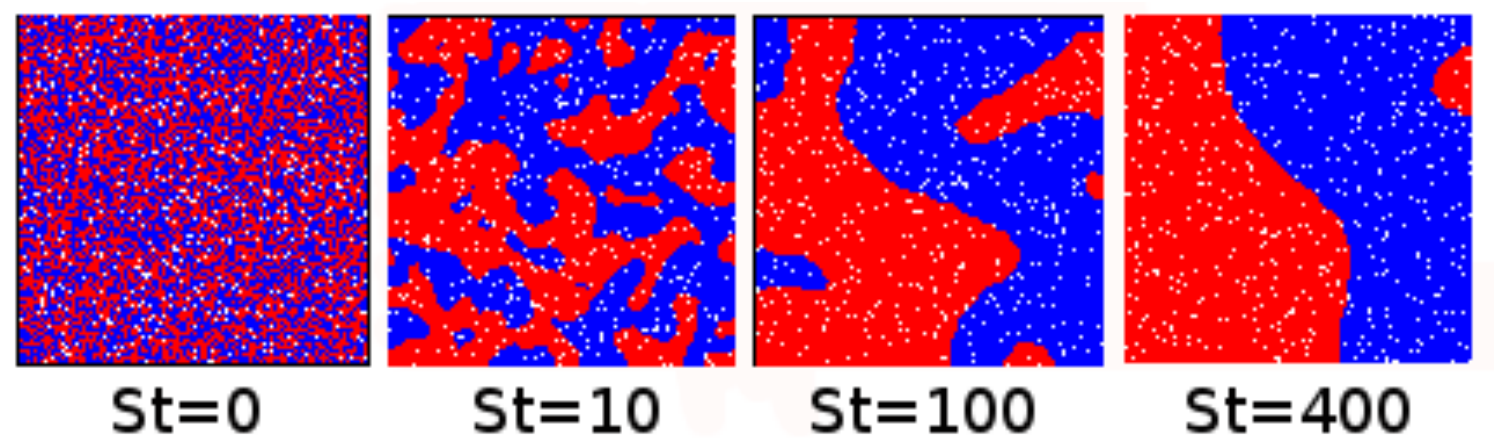}
\caption{Evolution of the configuration for a vacancy concentration $\rho=5\%$ and a tolerance $T=0.5$ with a network size $L=100$. $St$ stands for the number of time steps. The  red and blue pixels correspond  to the two types of agents, the white pixels to the vacancies. The system evolves from a random configuration -- where the vacancies and the two types of agents are intimately mixed -- to a completely segregated configuration. After just $10$ steps there exist two percolating clusters, one of each color, which are very convoluted, fractal-like.}

\label{exemple} 
 \end{center}
 \end{figure}
 
\begin{figure}[h]
 \begin{center}
 \includegraphics[width=8.7cm] {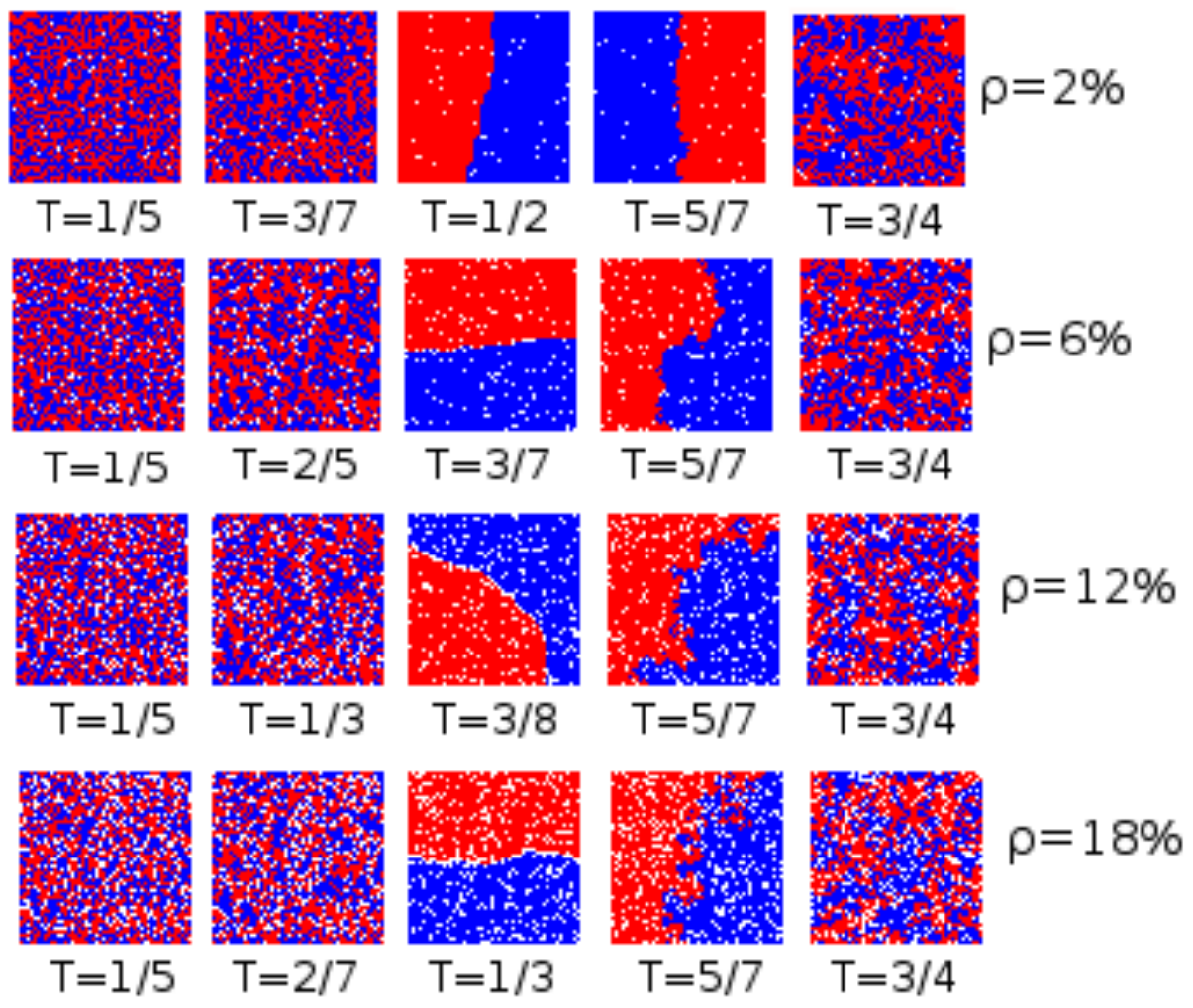}
\caption{Configurations obtained at large times for selected values of $\rho$ and $T$.}
\label{vois}
 \end{center}
 \end{figure}

Fig.~\ref{exemple} shows the time evolution of a typical configuration, for a vacancy density of $5\%$ and a tolerance $T=0.5$, which means that agents accept at most half of their neighbors to be different from themselves. 
One observes the rapid formation of large clusters. After $10$ time steps, the proportion of satisfied agents is already very close to $1$ (Appendix, Fig.~\ref{evolution}). It increases slowly thereafter, but the structure of the clusters keeps evolving.  They become more and more compact and well separated spatially, their surface gets less corrugated, in a process strongly reminiscent of coarsening effects in alloys  \cite{Bray}. We now consider more general values of the tolerance and the vacancy density. Fig.~\ref{vois} shows configurations obtained 
after letting the system evolve with the dynamics previously described until it reaches equilibrium. 
What is meant by equilibrium here may correspond to two different situations:  (i) The system does not evolve at all anymore (fixed point); (ii) The systems reaches some stationary state: the fluctuations of the studied parameters remain weak during a large number of time steps. In the following all averaged quantities are measured during $30000$ 
steps after equilibrium is reached.

At small and moderate values of $\rho$, one observes that :\\
\indent -- For low values of the tolerance the system stays in a mixed state, no large one-color clusters are formed although this would be more satisfactory for the agents. Actually, whatever the initial configuration, the system remains close to the state in which it was created: this is a dynamically frozen state.\\
\indent -- When $T$ increases, at fixed 
 $\rho$, a drastic qualitative change occurs at an intermediate value of $T$: The system separates into two homogeneous regions of different colors, segregation occurs.  
This behavior subsists for an interval of $T$ which depends on $\rho$.\\
\indent -- For $T$ larger than a value weakly  depending on  $\rho$,
 the final configuration is again mixed. \\ 
For large values of $\rho$ one observes a smooth transition as $T$ increases from 
a segregated state to a mixed one.
We will now give a more quantitative description and characterize the transitions between these different behaviors.

\section{Quantitative analysis: order parameters}

\subsection{ Main order parameter: A measure of segregation}
\begin{figure}[h]
\begin{center}
\includegraphics[width=10cm] {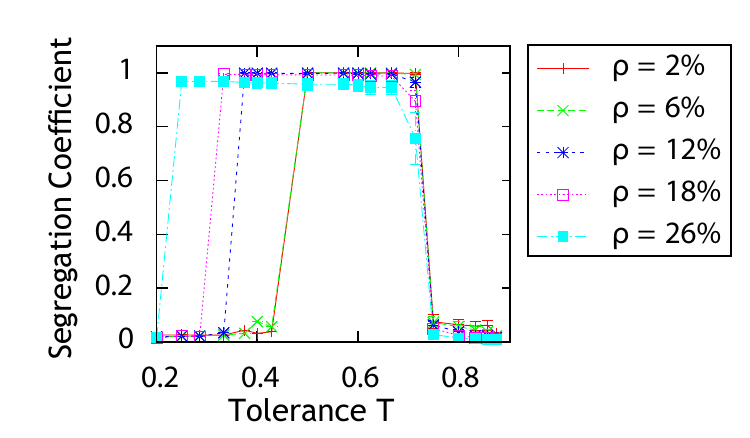}
\caption{Segregation coefficient (average of $s$ defined in Eq. (\ref{eq:s})) for several values of the vacancy density $\rho$.  
 The lines linking the points are guides to the eye.}
\label{seg}
\end{center}
\end{figure}

\begin{figure}[h]
\begin{minipage}[t]{.4\linewidth}
\begin{center}
 \includegraphics[width=3cm] {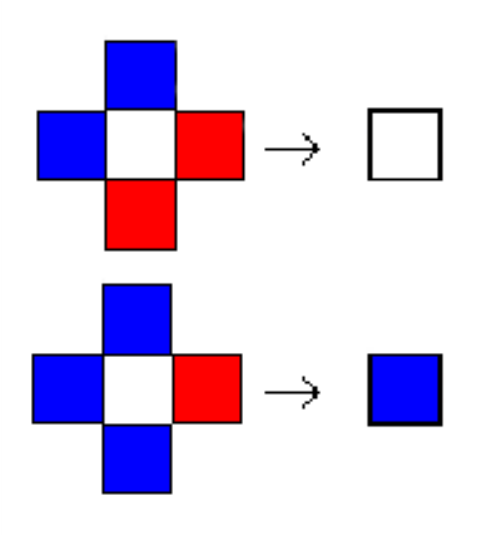}
\end{center}
\end{minipage}
\hfill
\begin{minipage}[t]{.5\linewidth}
\begin{center}
\includegraphics[width=4.5cm] {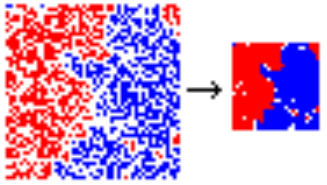}
\end{center}
\end{minipage}
\label{r}
\caption{Example of renormalization. The renormalization is performed on a configuration corresponding to $T=\frac{1}{5}$ and $\rho=50\%$.}
\end{figure}

\begin{figure}[h]
\begin{center}
\begin{minipage}[t]{.4\linewidth}
\includegraphics[width=6cm]{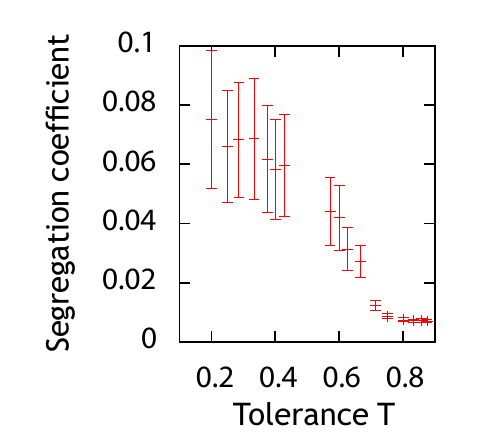}
\end{minipage}
\hfill
\begin{minipage}[t]{.5\linewidth}
\includegraphics[width=6cm]{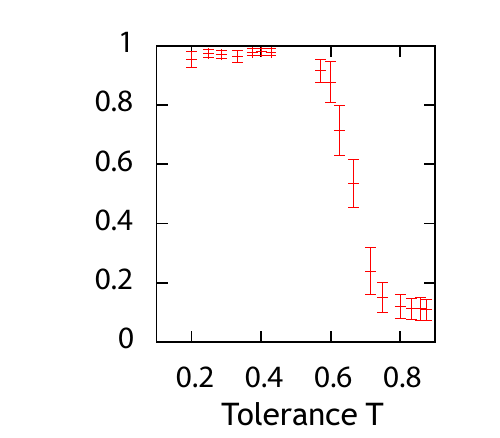}
\end{minipage}
\caption{Segregation coefficient as a function of the tolerance for $\rho=50\%$: Raw data  
(left) and results for the renormalized systems (right). These suggest that a continuous transition takes place near $T = 0.6 $. The error bars were obtained by computing the variance $\sigma_s$, $\sigma_s^2 = \langle s \rangle^2 - \langle s^2 \rangle$ (same for  Fig.\ref{seg}).}
\label{renorm50}
\end{center}
\end{figure}

Though the presence of segregation in a system can be visually assessed, a quantitative way to measure it is necessary, in particular to discuss the nature of the transitions between different states. Different possible measures have been suggested, by Schelling himself and later by various authors \cite{PancsVriend,SiVaWe,Fagiolo2}, which capture various aspects of the phenomenon.
Here we introduce a  measure linked to the  definition of segregation as the grouping of agents of the same type \emph{and} the exclusion of the other type in a given area. To that effect we consider that two agents belong to the same cluster if they are nearest neighbors (the Moore neighborhood cannot be used here, since Moore clusters of different colors may overlap and be large without segregation occurring). The (mass) size  of a cluster $c$, i.e the number of agents it contains, will be called $n_c$. Taking a hint from percolation theory where it plays a central role \cite{StAh}, we introduce the weighted average $S$ of the size of the clusters in one configuration
 \begin{equation}
 S=\sum_{\{c\}} n_c \; p_c 
\end{equation}
where $p_c=\frac{n_c}{N_{tot}}$ is the weight of cluster $c$, $N_{tot} = L^2(1-\rho)$ being the total number of agents. The maximal size of a cluster is $N_{tot}/2$, so the normalized weighted cluster size  $s$ is given by
\begin{equation}
 s =\frac{2}{L^2(1-\rho)} S = \frac{2}{{(L^2(1-\rho)})^2} \sum_{\{c\}} {n_c}^2
\label{eq:s}
\end{equation}
The sample average of $s$ after reaching equilibrium will be called the segregation coefficient $\langle s \rangle$. Its value for complete segregation (i.e., only two clusters survive) is $1$ and it vanishes if the  size of the clusters remains finite when the system dimension $L$ tends to infinity. It may therefore play the role of an order parameter to identify a segregation transition. The variation of the segregation coefficient $\langle s \rangle$  with respect to the tolerance is illustrated  in Fig.~\ref{seg} for different values of the vacancy density. The calculations were done using the Hoshen--Kopelman algorithm to labelize the clusters \cite{HK}.
For each density of vacancies there exist two critical values of the tolerance. At the first one $T_{f}$ ($f$ for frozen), $\langle s \rangle$ 
jumps from a very low value to about $1$. This signals an abrupt change from a mixed configuration to one with only two clusters (one for each type of agent). A second jump, in the reverse direction,  occurs for a larger tolerance $T_c$. This second value depends slightly on $\rho$, unlike $T_{f}$. The higher the density of vacancies, the smaller the value of $T$ at which the segregation phenomenon appears and the broader the interval of $T$ for which it exists. For a vacancy density above about $20 \%$ the segregation coefficient departs from $1$ (Fig.~\ref{seg}), even if the agents are visually segregated (Fig.~\ref{r}). This is due to the definition of a cluster based only on the four nearest neighbors. In some regions, even if there are agents of one color only the presence of many vacancies may lead to group them into distinct clusters and to miss the existence of a "dilute segregation" situation. In order to identify clusters at a larger scale in such cases we introduce a real-space renormalization \cite{rsrg} procedure. An example of the renormalization process for a site and its neighborhood is illustrated on the left of Fig.\ref{r}. A renormalized configuration is shown on the right part of  Fig.\ref{r} (see the appendix for details on the procedure).
The renormalization has a strong effect for high values of the density of vacancies. For $\rho=50\%$  the configurations are visually segregated for a range of small and medium values of the tolerance but the raw segregation coefficient is very small (Fig.~\ref{renorm50}, left), whereas after renormalization it is very close to $1$ (Fig.~\ref{renorm50}, right).
We can conclude that, for values of the vacancy density strictly less than $50\%$, there is a discontinuous transition from the segregated to the mixed state (as shown on Fig.~\ref{seg} for $\rho$ up to $26\%$). 
At $\rho=50\%$ the transition becomes continuous -- as shown on Fig.\ref{renorm50} the order parameter has a smoother variation --, and there is no longer any clear transition for vacancy densities
above $\sim 56\%$ (not shown). Moreover, the configurations observed at large $\rho$  
suggest the existence of a diluted phase of segregation at high vacancy densities:  
the two types of agents are not mixed but there are domains with many small clusters of a same color in a sea of vacancies. As for the frozen state observed at low $T$ for small $\rho$, it disappears at some medium value of $\rho$ (compare Fig.~\ref{seg} with Fig.~\ref{renorm50}, right). 

\subsection{A second order parameter: densities of unwanted locations.}
The segregation coefficient defined above does not allow to distinguish between the mixed state at low tolerance and the one at high tolerance (Fig.\ref{seg}). However, even if the final configuration is a mixed configuration for both situations, the nature of the two states is not the same. An additional parameter is thus necessary to analyze the results. We introduce the density $\tilde{\rho_{r}}$ of empty places where the red agents do not want to move  (symmetrically $ \tilde{\rho_b}$ for blue agents).  
\begin{figure}[h]
\begin{center}
\includegraphics[width=10cm]{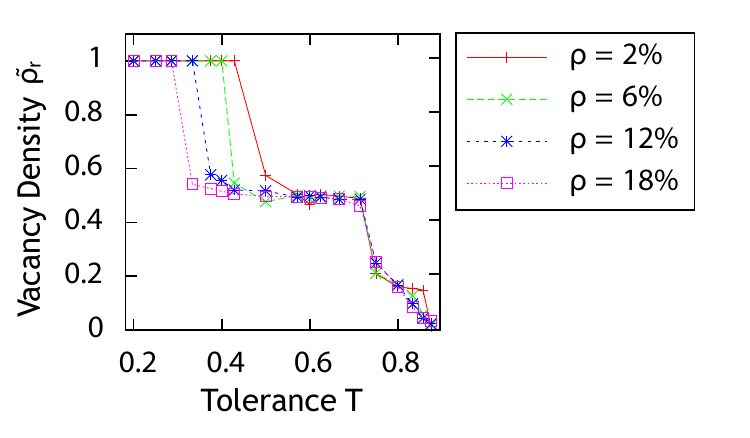}
\caption{Density $\tilde{\rho_{r}}$ of vacancies where the red agents would be unsatisfied, for several vacancy densities. The results for the blue agents are similar. Indeed, they play symmetrical roles in the present model.}
\label{lac}
\end{center}
\end{figure}
The plot of $\tilde{\rho_{r}}$ versus the tolerance (Fig.\ref{lac}) shows that this quantity  undergoes two jumps for each vacancy density $\rho$. Until the tolerance reaches $T_f$ no empty space is considered attractive by these agents.  Between $T_f$ and $T_c$ half of the vacant sites are satisfactory for one type of agents. For $T$  
larger than $T_c$ almost all empty spaces are acceptable. A representative configuration for a situation where all moves are permitted is an unordered one, this explains the mixed situation observed at high tolerances.
The quantity  $\tilde{\rho_{r}}$ thus allows  to characterize the three regimes and to discriminate between the low and high tolerance mixed situations.\\
 Thanks to the quantities defined previously we have identified the regions of segregation. Let us summarize the main results obtained so far.\
The final configurations are reminiscent of those encountered in spin lattice models with paramagnetic (= mixed state) and ferromagnetic (= segregated state)  phases.The segregation coefficient and the density of unilaterally unwanted locations play the role of order parameters. They show discontinuous jumps at some particular values of the tolerance $T$, indicating the existence of sharp phase transitions. For not too large values of the vacancy density there are two such transitions: At low tolerance, the system goes from a frozen state due to the dynamics that does not allow any movement, to a segregated state. At higher tolerance the system becomes mixed again.

\section{Analogues of thermodynamic quantities}

\subsection{ Contact with Spin-1 models and analogue of the energy.}
\begin{figure}[h]
\begin{center}
\includegraphics[width=10cm]{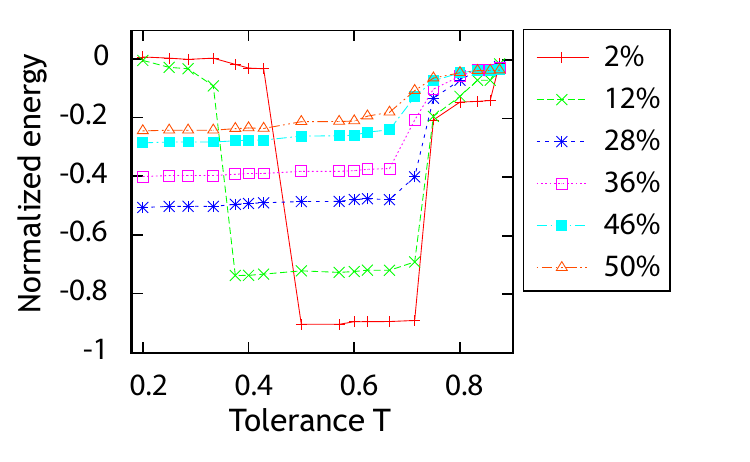}
\caption{Variation of the mean of the Blume Capel energy $E_{BC}$ for different values of the vacancy density. The data are normalized by $4L^2(1-\rho)$.}
\label{en}
\end{center}
\end{figure}

We now introduce several other useful quantities by analogy with thermodynamic properties studied in statistical physics. To do so, we first exhibit a link with spin-1 models. \\
One can show that (see the appendix and \cite{inprep}), had we forbidden the displacement of satisfied agents (as studied in \cite{schelling1978,Kirman,Marsili}), the dynamics would have a Lyapunov function, that is a quantity which decreases with time, driving the system towards a fixed point. This function can be written under the form 
\begin{equation}
 E_S= -\sum_{\langle i,j\rangle}c_ic_j -  K \sum_{\langle i,j\rangle} c_i^2 c_j^2 ,
\label{eq:Es}
\end{equation}
where  $K=2T-1$  and the $c_i$s are 'spin-1' variables taking the value $0$ if the location $i$ is not occupied and $1$ (resp. $-1$) if this location is occupied by a red (resp. blue) agent; the sums are performed on the nearest and next nearest neighbors. This function (\ref{eq:Es}) is identical to the energy of the Blume-Emery-Griffiths model \cite{BEG} under the constraint that the number of sites of each type ($0, \pm 1$) is kept fixed. This spin-1 model, and the Blume Capel model \cite{Blume,Capel} corresponding to the particular case $K=0$, have been used in particular to modelize binary mixtures and alloys in the presence of vacancies. A more detailed analysis of the link between the Schelling model and the Blume-Emery-Griffiths model will be published elsewhere \cite{inprep}. In the particular variant considered here, however, the dynamical rules do not lead to the minimization of  such a global energy. Yet, it is clearly potentially interesting to consider this quantity $E_S$  as a surrogate of the energy. Compared to the dynamics having $E_S$ as a Lyapunov function, the moves of satisfied agents introduce a source of noise which has some similarity with a thermal noise. Its amplitude may be measured by the fraction of agents who are satisfied. When starting from a random initial configuration this is higher at higher tolerance, hence the tolerance value is an indirect measure of this noise level. This gives another motivation, different from the one already evoked,  for taking the analogy between $T$ and a temperature as a guideline for the analysis, as done in what follows. We find that the average of the second part of the energy $E_S$ essentially consists of a term linear with $K$ (see appendix),
so that the transitions are more easily located by only plotting the average of the first term of $E_S$, Fig.~\ref{en}, corresponding to the Blume-Capel part of the energy, $E_{BC} = -\sum_{\langle i,j\rangle}c_ic_j$.
It confirms the existence of the two transitions previously evoked: At low tolerance its decrease occurs at the transition from the frozen state to the segregated one, whereas the increase observed at high tolerances corresponds to the transition to the mixed state. Such abrupt variations are characteristic of a discontinuous -- in thermodynamic language "first order" -- transition. Moreover we note from Fig.~\ref{en} that as the vacancy density increases the energy varies less abruptly: This signals a change in the nature of the transition, from discontinuous to continuous ('second order'), as will be discussed below.
\subsection{ Analogue of the specific heat}
Since the energy analogue  proved fruitful, an analogue of the specific heat may also be expected to give  useful information. However, here we cannot make use of the thermodynamic definition $C_s = dE_S/dT$ since only a finite number of values of the tolerance have physical meaning. In order to work with a well-defined quantity at fixed tolerance $T$  we remark that the specific heat is related to the energy fluctuations at equilibrium via the so-called fluctuation-dissipation theorem. In the present context the relevant formulation of this important theorem is:
\begin{equation}
 C_s =\frac{ \langle E_S^2 \rangle - {\langle E_S \rangle}^2}{T^2} 
\label{eq:cs}
\end{equation}
where  the notation $\langle \; \rangle$ means an average over the configurations taken by the system after reaching equibrium, and $T$ is the tolerance.
\begin{figure}[h]
\begin{center}
\includegraphics[width=10cm]{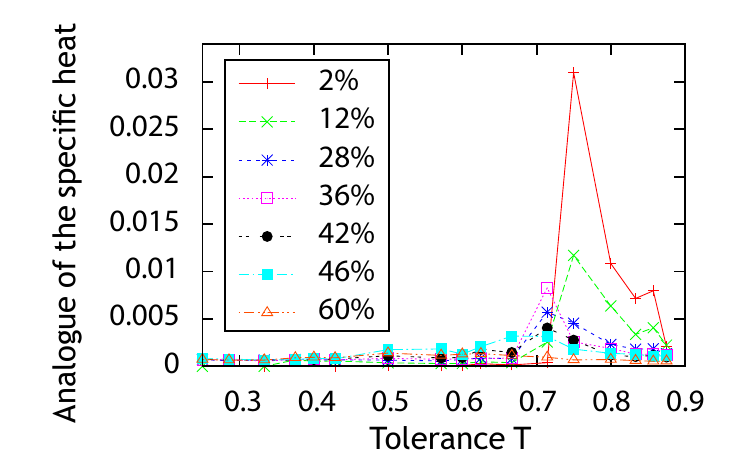}
\caption{Variation with the vacancy density of the fluctuation coefficient $C_s$.}
\label{spe_heat}
\end{center}
\end{figure}
 \noindent We will call $C_s$ the {\it fluctuation coefficient}. It plays a role analogous to the volatility index measuring price fluctations in financial markets. \\
According to Fig.~\ref{spe_heat} the fluctuation coefficient  has a well-marked peak at the ``segregated-mixed'' transition. This peak flattens out and tends  to disappear as the vacancy density increases, confirming  the disappearance of the transition from segregated to mixed states. On the other hand this fluctuation coefficient,
null at very low tolerance, has a slight increase at the ``frozen-segregated'' transition. The jump in the energy $E_S$ does not show up in the fluctuation coefficient $C_s$, indicating that the fluctuation-dissipation theorem is stongly violated, as typically observed in dynamic transitions in glassy or kinetically constrained systems \cite{RS}.

\subsection{ Analogue of the susceptibility}
\begin{figure}[h]
\begin{center}
\includegraphics[width=10cm]{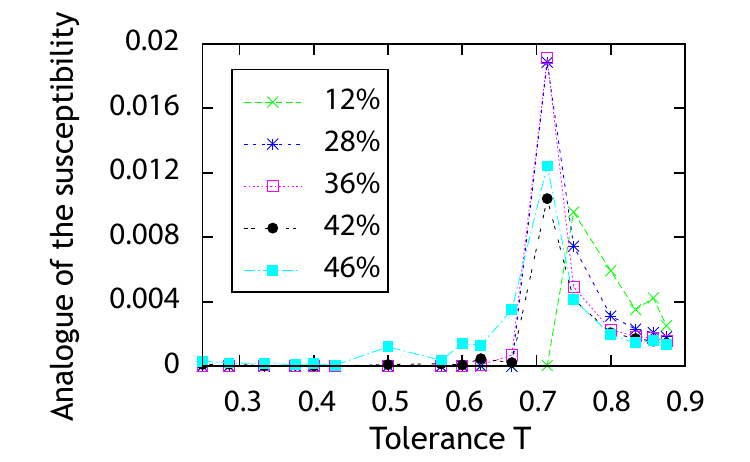}
\caption{Evolution with the vacancy density of $\chi_s $, analogue of a susceptibility.}
\label{suscept}
\end{center}
\end{figure}

In the same vein one would like to define the susceptibility, a quantity which is to the order-parameter what the specific heat is to the energy. Here, having proposed the segregation coefficient as the main order parameter, we define the analogue of the susceptibility through the fluctuation-dissipation relation involving the order parameter fluctuations
\begin{equation}
\chi_s =\frac{ \langle s^2 \rangle - {\langle s \rangle}^2}{T}
\label{eq:chi}
\end{equation}
where  $s$ is defined by Eq. (\ref{eq:s}). As can be seen on Fig.~\ref{suscept} this susceptibility presents a peak at the second transition which is more conspicuous for high vacancy densities than the one for $C_s$.
This peak does not exist anymore for high vacancy densities in agreement with the disappearance of the transition evoked in the previous part.

\section{Phase diagram}
\subsection{ Representation of the phase diagram}
\begin{figure}[h]
 \begin{center}
 \includegraphics[width=11cm] {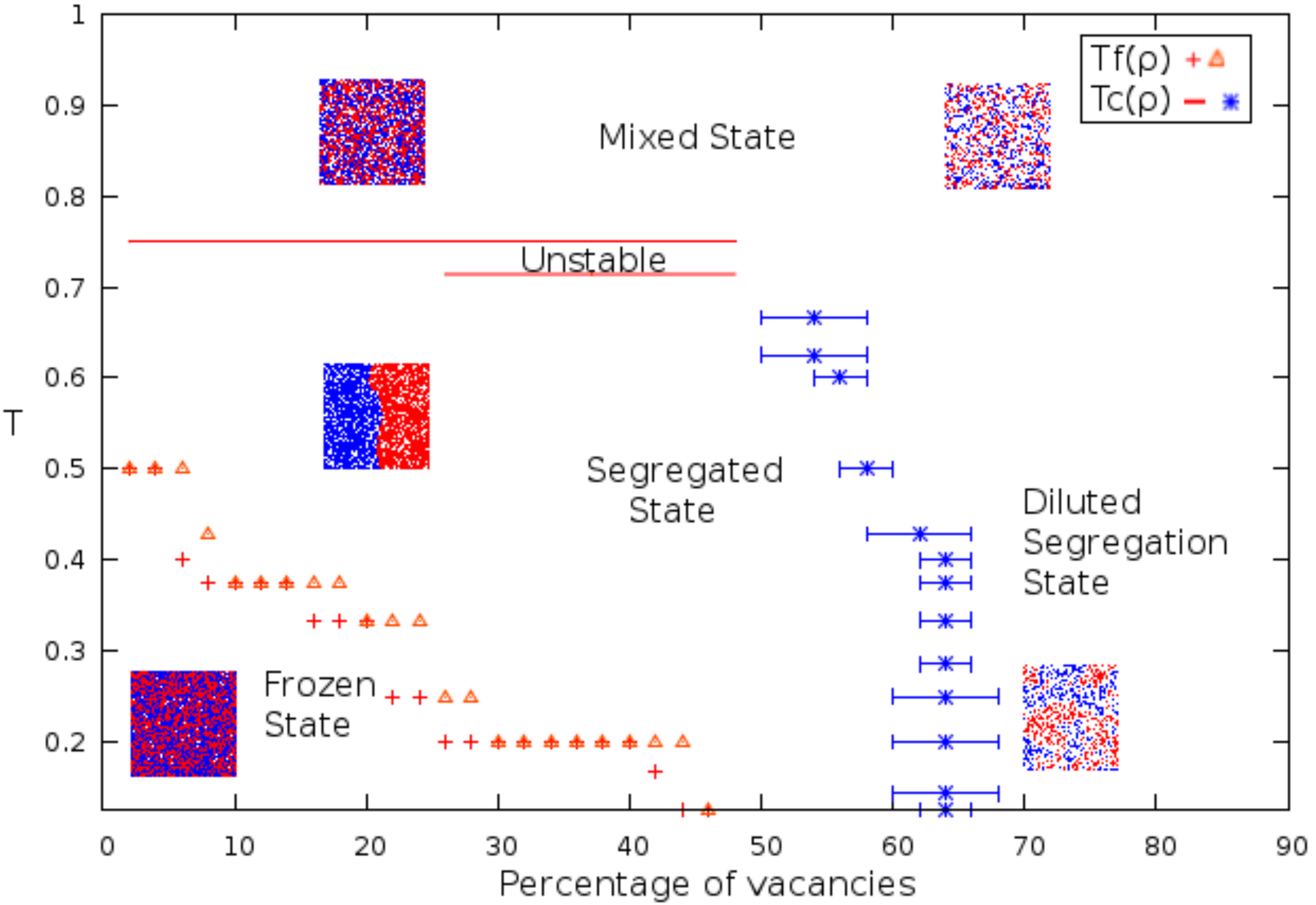}
 \caption{Phase diagram of the studied Schelling model. The blue crosses correspond to the continuous transition between the segregated and dilute segregated states. The red triangles and pluses are the  upper and lower  limits of the transition between the frozen and segregated states. The red lines separate the segregated state from the mixed one. Note that the tolerance $T$ only takes discrete values.}
\label{diag}
 \end{center}
 \end{figure}
We are now in position to build the full $(\rho, T)$ phase diagram. We have seen that the system can end up in four possible  states: (i) segregated state, (ii) mixed state, (iii) diluted segregation state,
(iv) frozen state. The domains of existence of these 'phases',
represented in the phase diagram, Fig.~\ref{diag}, can be briefly
described as follows: (1) For vacancy densities below $46\%$ a dynamical transition line $T_f(\rho)$ separates the frozen state from the segregated one. (2) Above this line the segregated state exists in a large domain bounded
by a transition line $T_c(\rho)$. (2a) This line is almost at a constant value of the tolerance for $\rho \lesssim  50\%$, and of a discontinuous nature (first-order like) up to  $\rho = 50\%$; it separates the segregated state from the mixed state. (2b) At $\rho= 50\%$, the transition becomes continuous, and the line becomes almost parallel to the $T$ axis; there it separates the segregated state from a diluted segregated one. (3) Beyond $T_c(\rho)$, at high values of the vacancy density one goes gradually (no sharp transition) from the diluted segregated state at low $T$ to the mixed one at high tolerance values. We now discuss in more detail the various phases and transitions involved.

\subsection{ Transition frozen state / segregated state} 

\begin{figure}[h]
\begin{center}
\includegraphics[width=10cm]{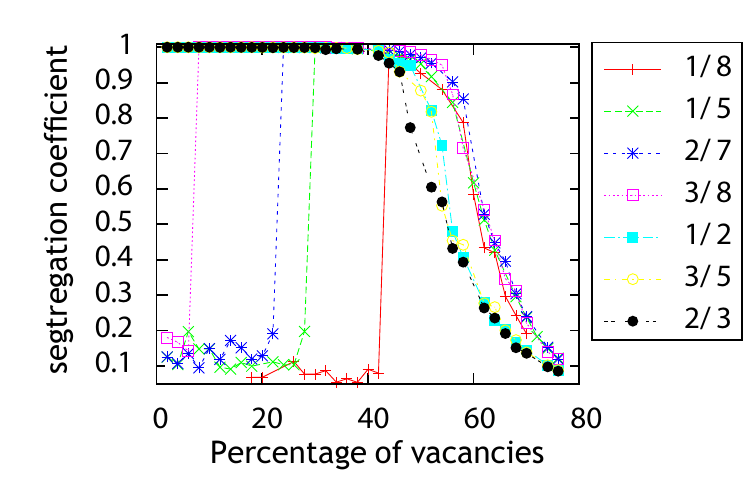}
\caption{Segregation coefficient for different values of the tolerance versus the vacancy density. }
\label{seg_den}
\end{center}
\end{figure}

At low tolerance there is a transition where the system abruptly switches from a frozen to a segregated state. The analogues of the specific heat and of the susceptibility  do not have a singular behavior in the vicinity of the transition (Fig. \ref{spe_heat} and \ref{suscept}). As we have seen, this is explained by the dynamical nature of this transition -- in contrast to the other transitions which are thermodynamical-like. We locate the transition by looking at the jump of the segregation coefficient, both at fixed $\rho$, increasing $T$ (Fig.~\ref{seg}), and at fixed $T$, increasing $\rho$ (Fig.~\ref{seg_den}). As the density of vacancies $\rho$ increases, the transition occurs for lower values of the tolerance. When this density is higher than $46\%$ this line of transition does not exist anymore.
We remark that this frozen-segregated state transition line can fluctuate depending on the order in which the agents are chosen during the dynamics: in an interval of tolerance and for a given vacancy density, the system may end either in a frozen state or in a segregated one. The corresponding range of tolerance is given in the appendix, Table \ref{tab}, for each tested vacancy density. Let us notice that inside this frozen phase, any initial configuration, segregated or not, with randomly distributed vacancies is very close to a stationary state.

\subsection{ Transition segregated state/mixed state.}
The segregated phase is upper bounded by a transition line  where the clusters disappear and the two types of agents become mixed.\\
\indent -- This change is abrupt for vacancy densities $\rho$ strictly smaller than $26\%$. Indeed, the plots of the segregation coefficient against the tolerance for values of $\rho$  in this range show a jump from $\sim 1$ to $\sim 0$ (Fig.~\ref{seg}).\\
\indent -- For  $26\%  \le \rho \le 48\%$, the mixed state is reached after an intermediate state where the system has no dynamical stability: it can oscillate between several acceptable configurations, leading to large fluctuations in the segregation coefficient.\\
\indent -- For a small range of values, $50\% \le \rho \lesssim 56\%$, the segregated states continously become mixed states. There, the line
abruptly turns downward. This area of the phase diagram, with a change in the nature of the transition and a sudden downturn of the transition line, is more difficult to study because of the discreteness of $T$ values and possible finite size effects.\\ 
For a given $\rho$ the value of the tolerance at which the segregated-mixed transition occurs can be located from the position of the peak of the analogue of either the specific heat or of the susceptibility (Fig.~\ref{spe_heat} and \ref{suscept}). In order to complete the study of this transition, we have considered the distribution of the segregation coefficient for several values of $T$ around the transition at different vacancy densities (see appendix). This illustrates the three ways to go from a segregated to a mixed state .\\

\subsection{ Transition segregated state/diluted segregation state.}
As just mentioned, the boundary $T_c(\rho)$ of the segregated phase becomes weakly dependent on the vacancy density $\rho$ when $\rho$ becomes  slightly larger that $50 \%$. To locate the transition line we plot for several values of $T$ the variation  with  $\rho$ of the segregation coefficient $C_s$ (Fig.~\ref{seg_den}), and  of the analogue of the susceptibility $\chi_s$ (see appendix Fig.~\ref{suscept_den}). At high vacancy densities, beyond the transition line, the nature of the phase is different at high and low tolerances. As we have seen, at high tolerance values there is a mixed state, whereas at low tolerance one finds a diluted segregation state that leads to very small segregation coefficients. Indeed, at high $\rho$ ($>0.592746$, the percolation threshold \cite{StAh}), the high probability of percolation of vacancies prevents the forming of large clusters. In this high vacancy density domain, we do not find any sharp transition from the diluted segregated state to the mixed one, but only a gradual change as the tolerance increases.

\subsection{ Comparison with the Blume-Capel phase diagram.}
  
It is instructive to compare the phase diagram with the one of the Blume-Capel model \cite{Blume,Capel} evoked above (see appendix). For this model a transition line separates a ferromagnetic (segregated) phase from a domain where one goes gradually from a paramagnetic (mixed) phase to a phase where the vacancies predominate. This line changes as well its nature from discontinuous to continuous. However, it is the ferromagnetic-paramagnetic transition which is second order, whereas the segregated-mixed transition is first order-like (the change is abrupt). Conversely, the transition between the ferromagnetic and the high vacancy density phases is first order, whereas the corresponding transition in the Schelling model is continuous.

\section*{Conclusion}
We analyzed a variant of the Schelling model from a physical point of view. We have introduced a measure of segregation and analogues of physical quantities such that the fluctuation coefficient and the susceptibility 
where, remarkably, the analogy between the tolerance and the temperature  proves fruitful. These quantities allowed to identify the different phases of the system and characterize the transitions between them (thermodynamical or dynamical like, discontinuous or continuous). The main results have been summarized as a phase diagram in the $(\rho, T)$-plane where $\rho$ is the vacancy density and $T$ the tolerance. Considering larger neighborhoods would allow to have a larger set of values for $T$ and approach a continuous model. A more precise location of the phase boundaries, if needed, would require computationally costly simulations of larger network sizes.\\
We have seen in particular  that the segregated phase occupies a large domain (up to a tolerance $T$ as high as $3/4$),
confirming Schelling's intuition on the genericity of the segregation phenomenon. The abrupt transition from a mixed to a segregated state could be interpreted as the tipping point -- more precisely the rapid ethnic turnover -- observed and studied by social scientists \cite{Neighborhood}. Besides, the diluted segregation state might be relevant for low-density suburban areas. The frozen state would probably be unstable in a more realistic model allowing for migratory flows of discontent agents to other cities.\\
The tools and methods presented here could be used to study other Schelling-like models.
Clearly future works should focus on models grounded on empirical data
and where the agent decision rules take into account relevant socio-economic factors  \cite{Clark}. Yet it is known from a large body of work in statistical physics that one
needs also to explore more widely the space of models in order to
identify what makes a particular behavior specific or generic.
As already mentioned our goal here was
to provide generic tools for the analysis of models of socio-dynamics.
The variant of Schelling's segregation model we have studied as a test of our approach has the advantage of being identical or very close to variants already studied in the literature and to allow links with known - and non trivial - spin models.

\section*{Acknowledgements}
LG is supported by a fellowship from the French Minist\`ere de l'Enseignement Sup\'erieur et de la Recherche, allocated by the {\em Ecole doctorale de Physique de l'UPMC -- ED 389}. JV and JPN are CNRS members.
This work is part of the project 'DyXi' supported by
the  SYSCOMM program of the French National Research Agency (grant ANR-08-SYSC-008).


\newpage

\appendix
\makeatletter
\renewcommand{\thefigure}{\ifnum \c@section>\z@ \thesection.\fi
 \@arabic\c@figure}
\@addtoreset{figure}{section}
\makeatother

\section{Appendix} 
$\;$\\
\subsection{Numerical simulations}
$\;$\\
All the simulations except on the Fig.~\ref{exemple} were performed on a $50 \times 50$ lattice. We tested all the meaningful values of the tolerance $T$ at even values of the vacancy percentage $\rho$. With this choice for $\rho$, the number of vacancies and of agents of two colors are  exactly equal to the integers $\rho*L^2$ and $L^2(1-\rho)$. To take only even values of the vacancy percentage $\rho$ also allows to moderate the computational cost. 
$\;$\\
\subsection{Real-Space Renormalization procedure}
$\;$\\

To identify clusters at a larger scale, we performed the following renormalization procedure. We divide the lattice into squares of $4$ sites. On each of these little squares, we look at the bottom right site :\\ 
\indent -- If this site and its neighborhood comprise a majority of blue (resp. red) agents, the $2 \times 2$ square is
replaced by a (single) blue (resp. red) agent.\\
\indent -- If this site and its neighborhood consist of a majority of vacancies, the $2 \times 2$ square is replaced by a vacancy.\\
\indent -- If there is no majority, the $4$-site square is replaced by an agent of the same type as the bottom right site (or by a vacancy if that site is empty).

\subsection{Contact with spin-1 models}
\subsubsection{{\it Energy of Schelling models}}
For completeness we present here a correspondence, discussed elsewhere \cite{inprep},  between Schelling segregation models and spin-1 models.\\ 
One can associate to each site $i$ of the lattice a spin variable $c_i$, taking the value $0$ if the location is not occupied, and $1$, resp. $-1$, for red, resp. blue, occupied sites. With these
'spin-1' variables the satisfaction condition for location $i$ (including the case where $i$ is vacant) can be written as:
\begin{equation}
 c_i \sum_{j\in (i)}  c_j + (2T-1)  c_i^2 \sum_{j \in(i)} c_j^2  \ge 0
\label{SI-eq:Sat}
\end{equation}
where $j\in (i)$ means $j$ belonging to the neighborhood of site $i$.
This suggests to define, as an analogue of the energy, 
\begin{equation}
 E_S= -\sum_{\langle i,j\rangle}c_ic_j -  K \sum_{\langle i,j\rangle} c_i^2 c_j^2 ,
\label{SI-eq:Es}
\end{equation}
where  $K=2T-1$ ($-1 \le K \le 1$), and the index $S$ stands for ``Schelling''.\\
For the Schelling original model, as well as for other variants where only unsatisfied agents can move, one can show \cite{inprep} that the energy $E_S$ is indeed a Lyapunov function, that is a quantity which decreases with time during the dynamics, driving the system towards a fixed point.
Note that the energy is not proportional to the global utility $U=\sum_i u_i$, where $u_i$ is $1$ if agent $i$ is satisfied, and $0$ otherwise. \\

This function $E_S$, (\ref{eq:Es}), is identical to the energy of the Blume-Emery-Griffiths model \cite{BEG} under the constraint that the number of sites of each type ($0, \pm 1$) is kept fixed. This spin-1 model, and the Blume Capel model \cite{Blume,Capel} corresponding to the particular case $K=0$, have been used in particular to modelize binary mixtures and alloys in the presence of vacancies. In the standard versions of these models, the energy contains the additional term $ D \sum_i {c_i}^2 $ (the sum being over all the sites), so that the total number of vacancies is fixed only in average through the Lagrange multiplier $D$:
\begin{equation}
 E_{BEG}= -\sum_{\langle i,j\rangle}c_ic_j -  K \sum_{\langle i,j\rangle} c_i^2 c_j^2 + D \sum_i {c_i}^2
\label{SI-eq:EBEG}
\end{equation}
The limit $D \rightarrow  - \infty$ corresponds to the absence of vacancies, i.e. the Ising model.
Large positive $D$ corresponds to high vacancy densities. The  term $D$ does not appear in the energy of the Schelling  model, not because it corresponds to $D = 0$, but  because the density of vacancies is fixed. The fact that $E_S$ is a Lyapunov function for the Schelling model where only unsatisfied agents move, means that such a model is equivalent to a Blume-Emery-Griffiths model without thermal noise (zero temperature), and under kinetic constraints (e. g., no direct exchange between two agents of different colors is allowed).\\

The standard order parameters for these spin-1 systems are the magnetization $(1/N) \sum_i c_i$
and the quadrupole moment, $(1/N) \sum_i c_i^2$, but here these quantities which are, respectively, the difference between the total numbers of agents of different colors, and the density of occupied sites,
are kept fixed by construction in the Schelling model. .

\subsubsection{{\it Blume-Capel model: phase diagram}}
\begin{figure*}[h!]
\begin{center}
\includegraphics[width=10cm]{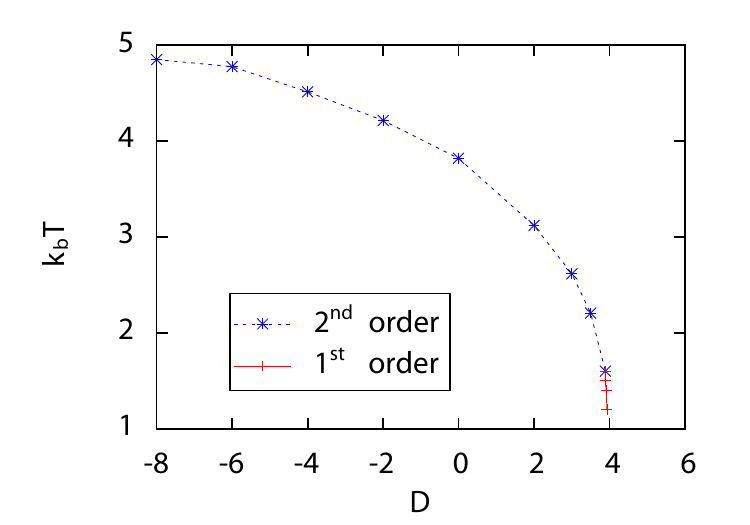}
\caption{Phase Diagram of the Blume Capel model with first and second neighours interactions (Moore neighborhood). The simulations were performed using the Heat Bath algorithm.}
\label{phasediag}
\end{center}
\end{figure*}

A striking similarity exists between the phase diagram in the $(\rho,T)$ plane of the variant of the Schelling model studied here, and the one of the Blume-Capel model \cite{Blume,Capel} in the $(D,T)$ plane -- where $T$ is the temperature and $D$ the parameter fixing in average the vacancy density.\\ 
The Blume-Capel phase diagram, computed for the Moore neighborhood, is shown on Fig~\ref{phasediag},
and should be compared with the one on Fig~\ref{diag}. The red part of the transition line corresponds to a first-order (discontinuous) transition, and the blue part to a second-order (continuous) transition. Below the transition line the system is in an ordered, ferromagnetic, state. Above the transition line, at low and medium values of $D$, one finds the unordered, paramagnetic, phase, and at large $D$ and low temperature, a vacancy dominated phase. In the latter phase, the typical configurations are not strictly comparable to the ones of the corresponding phase in the Schelling model: in the spins models, the clusters are compact whereas it is not the case in the Schelling model. In the Blume-Capel model there is no frozen phase, because there is no constraint on the dynamics.

$ \;$
$ \;$

The transition line in the Blume-Capel model, shown on Fig~\ref{phasediag}, has been built up by plotting for fixed temperature $T$ (resp. fixed $D$ depending on the area of the diagram dealt with) the magnetization versus $D$ (resp. $T$) for three different sizes, and by looking at the value of $D$ (resp. $T$) corresponding to the intersection of the three curves.

\newpage
\subsection{Complementary analysis of the segregation model}

\subsubsection{{\it Density of satisfied agents}}
\begin{figure*}[h]
\begin{center}
  \includegraphics[width=8cm]{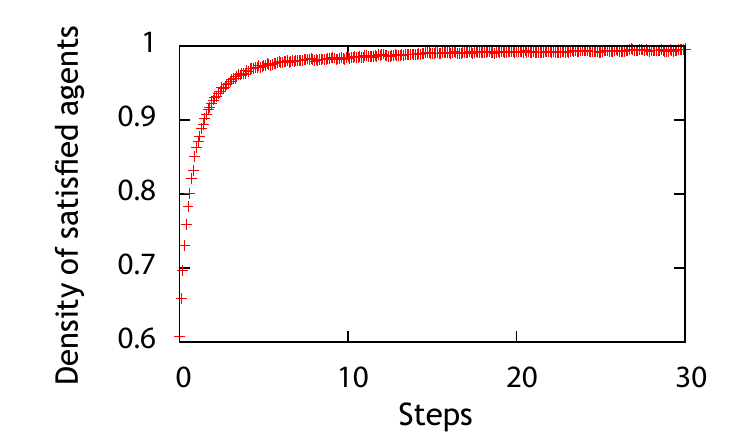}
\caption{Evolution of the proportion of satisfied agents versus the number of steps for $\rho=5\%$ and $T=0.5$. About $60\%$ of the agents are initially satisfied.  The dynamics quickly allows the agents to be almost all satisfied, after only $10$ steps the density of satisfied agents is very close to $1$. It increases slowly afterwards, with small fluctuations due to the possibility for satisfied agents to keep moving.}
\label{evolution}
\end{center}
 \end{figure*}

The density of satisfied agents is a good indicator of the convergence of the system towards its stationary state (whenever it exists), in which the fraction of satisfied agents is constant. Fig.~\ref{evolution} shows the evolution of the density of satisfied agents for a case with low vacancy density $\rho=5\%$ and moderate tolerance $T=0.5$. We note that, in the unstable part of the phase diagram (close to $T_c$ at $26\% \leq \rho \lesssim 50\%$), the density of satisfied agents is stationary: this shows that, more generally, the convergence of the fraction of satisfied agents does not guarantee that the system itself has reached a steady state.

\subsubsection{{\it Unwanted vacancies}}
\begin{figure*}[ht]
\begin{center}
\includegraphics[width=10cm]{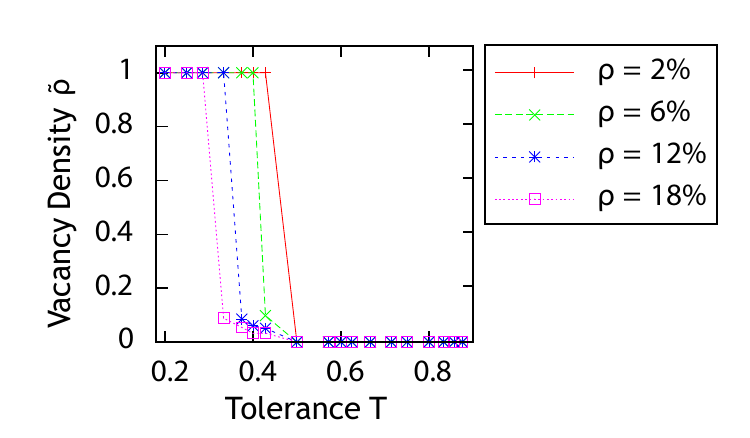}
\caption{Density $\tilde{\rho}$ of vacancies where no type of agent would be satisfied, for several values of the vacancy density. }
\label{lac2}
\end{center}
\end{figure*}

The variation with the tolerance of the unilaterally unwanted density of vacancies $\tilde{\rho}$  (Fig.\ref{lac2}) confirms that the mixed situation  observed for low values of  $T$ is due to the dynamics.
The agents reject all the empty spaces, consequently the system cannot evolve. 
At tolerances corresponding to the frozen state - segregated state transition, the situation reverses. All vacancies are acceptable for at least one type of agents.\\

\subsubsection{{\it Analogue of the susceptibility}}
The analogue of the susceptibility for the Schelling model is given by:
\begin{equation}
\chi_s =\frac{ \langle s^2 \rangle - {\langle s \rangle}^2}{T},
\label{eq:si-chi}
\end{equation}
where  the notation $\langle \; \rangle$ means an average over the configurations taken by the system after reaching equibrium, and
$T$ is the tolerance.
\begin{figure*}[h]
\begin{center}
\includegraphics[width=10cm]{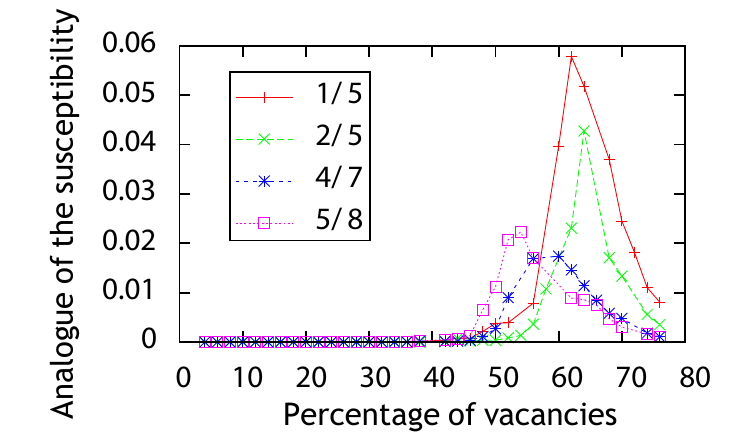}
\caption{Analogue of the susceptibility for different values of the tolerance versus the vacancy density. The averages have been computed on $30000$ simulations after equilibrium.}
\label{suscept_den}
\end{center}
\end{figure*}
These fluctuations allow to locate the transition at fixed tolerance $T$ as the density of vacancies increases.

\subsubsection{{\it Blume-Emmery-Griffiths energy vs Blume-Capel energy}}
\begin{figure*}[h]
\begin{minipage}[t]{.48\linewidth}
\begin{center}
\includegraphics[width=7cm]{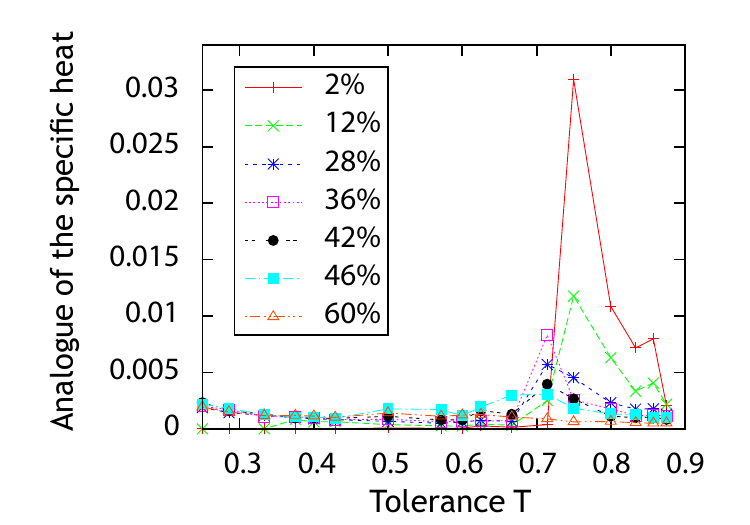}
\caption{Variation of the analogue of the specific heat obtained from the Blume-Capel energy for different values of the vacancy density.}
\label{spe_heat_BC}
\end{center}
\end{minipage}
\hfill
\begin{minipage}[t]{.48\linewidth}
\begin{center}
\includegraphics[width=7cm]{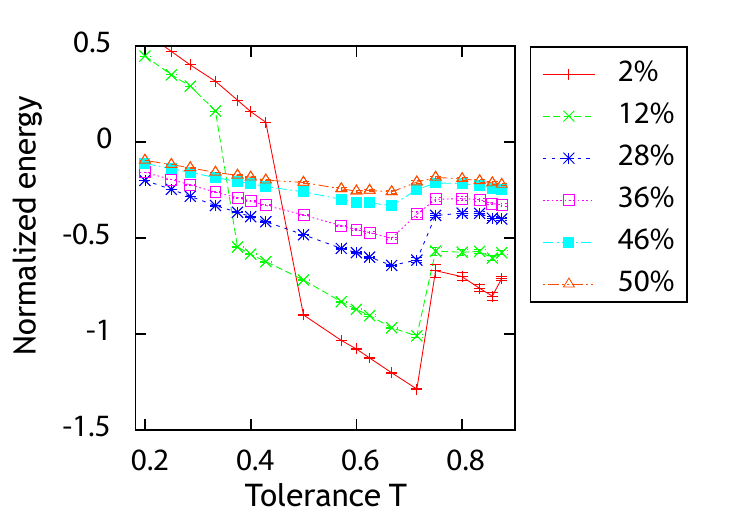}
\caption{Variation of the mean of the  energy $E_S$  for different values of the vacancy density. The data are normalized by $4L^2(1-\rho)$.}
\label{en2bis}
\end{center}
\end{minipage}
\end{figure*}

We have shown that the appropriate energy related to the Schelling model is the 
Blume-Emmery-Griffiths energy at constant number of sites of each type (red and blue agents, and vacancies), defined by (\ref{SI-eq:Es}). However we find that the analysis of the transitions can be done as well from the Blume-Capel energy, that is here, since the number of sites of a given type is fixed (no $D$-term):
\begin{equation}
 E_{BC}= -\sum_{\langle i,j\rangle}c_ic_j
\label{SI-eq:EBC}
\end{equation}
Note that in the absence of vacancies, 
$c_i=\pm 1$ so that $E_{BC}$ would reduce to the standard Ising energy.
One finds that the fluctuation coefficient obtained with the Blume-Capel energy, defined by $C_s' =\frac{ \langle E_{BC}^2 \rangle - {\langle E_{BC} \rangle}^2}{T^2}$ and  shown on Fig~\ref{spe_heat_BC}, is very similar to the one obtained from the Schelling energy $E_S$. Actually, in the evolution of the mean of the total energy $E_S$, shown on Fig.~\ref{en2bis}, one recognizes the contribution from the Blume-Capel energy $E_{BC}$ (shown on Fig.~\ref{en}),
simply increased by an additional term linear in $K=2T-1$. These observations can be explained by the fact that the difference between these two energies is proportionnal to the numbers of pairs of occupied sites of which the fluctuations are weak. 
Indeed, one observes that the vacancies remain approximatively uniformly distributed. This comes from the fact that each agent searches for locations where the number of unlike neighbors is inferior to a given {\em proportion} of the number of neighbours. Note that the vacancies would not be uniformly placed if each agent wanted less than a fixed {\em number} of unlike neighbors, as considered in \cite{SiVaWe}.

\subsubsection{{\it Segregation coefficient}}
\begin{figure*}[ht]
\begin{minipage}[t]{.50\linewidth}
\begin{center}
\includegraphics[width=7cm]{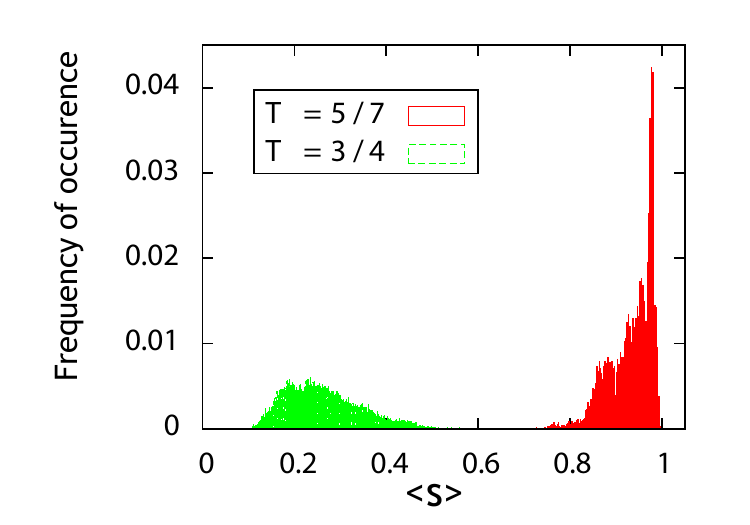}
\end{center}
\end{minipage}
\hfill
\begin{minipage}[t]{.50\linewidth}
\begin{center}
\includegraphics[width=7cm]{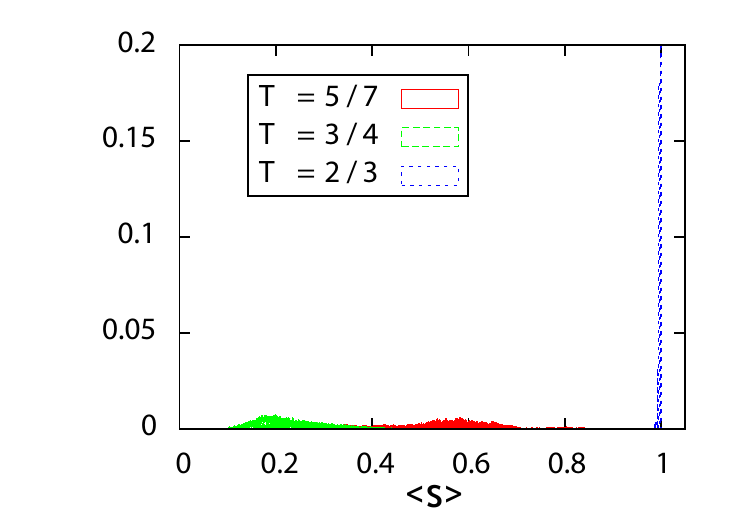}
\end{center}
\end{minipage}
\caption{Distributions (normalized by the number of measures) of the segregation coefficient for  $\rho=24\%$ and $\rho=28\%$.  }
\label{seg_dist28}
\end{figure*}

\begin{figure*}[ht]
\begin{minipage}[t]{.50\linewidth}
\begin{center}
\includegraphics[width=7cm]{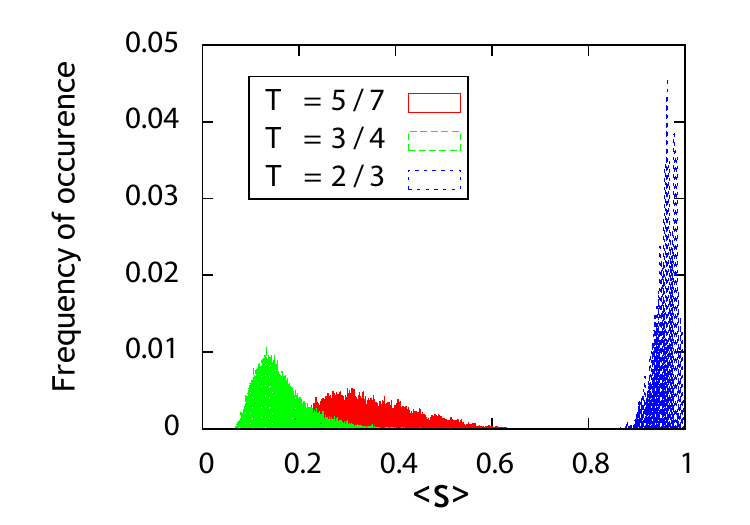}
\end{center}
\end{minipage}
\hfill
\begin{minipage}[t]{.50\linewidth}
\begin{center}
\includegraphics[width=7cm]{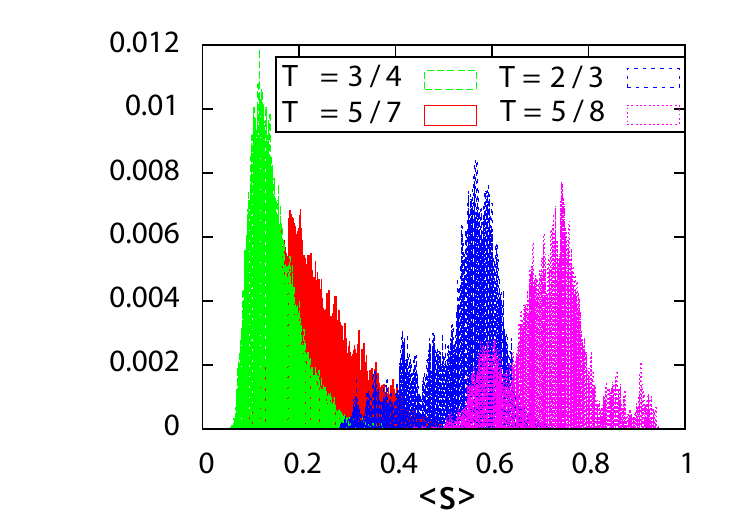}
\end{center}
\end{minipage}
\caption{Distribution of the segregation coefficient for $\rho=44\%$ and $\rho=50\%$.}
\label{seg_dist50}
\end{figure*}

In order to get more information about the segregated state-mixed state transition, it is instructive to look at the distribution of the segregation coefficient $<s>$ in the vicinity of the transition (Figs \ref{seg_dist28} and \ref{seg_dist50}). The distributions have been obtained from $30000$ calculations of the segregation coefficient.\\
At vacancy density  $\rho$ lower than $26\%$, the transition is well marked. This distribution is centered near $0$ at the transition point whereas it is centered near $1$ at the closest inferior value of the tolerance. Indeed, the distributions of the segregation coefficient  for successive tolerances around the transition are clearly separated for the vacancy density $\rho=24\%$ (left Fig.~\ref{seg_dist28}).\\
As $\rho$  increases, the transition is achieved via an intermediate state. One distinguishes three kinds of distributions (Figs.~\ref{seg_dist28}, right and  \ref{seg_dist50}, left) which are centered around a very small value ($\sim 0.2$), peaked close to $1$, or centered around an intermediate value. The latter case corresponds to a broader distribution.

Once the vacancy density is greater than $50\%$, all the distributions of the segregation coefficient begin to blend together (see Fig \ref{seg_dist50}, right). This confirms that the increase in the vacancy density is accompanied by a change in the nature of the transition between $46\%$ and $50\%$, which becomes continuous.

\newpage
\subsubsection{{\it ``Frozen-segregated'' transition line}}

The transition line between the frozen and the segregated states has been determined by locating the jump of the segregation coefficient $<s>$ from $\sim 0$ to $\sim 1$. The initial configuration and the order of choice of the agents, when the dynamics is applied, create fluctuations on the limit between the two states. Actually, for some sets of parameters $(\rho,T)$, the system may end either in blocked or in segregated configurations depending on the order of choices of the agents. One cannot exclude that this unstable domain is due to finite size effects, hence disappearing in the infinite network size limit. However, it is not surprising to find such a metastability effect close to a discontinuous transition.

 \begin{table}[ht]
\begin{center}
\begin{tabular}{@{\vrule height 5.5pt depth3pt  width0pt} p{0.2cm} p{0.7cm} p{0.5cm} p{0.7cm} p{0.7cm} p{0.7cm} p{0.7cm} p{0.5cm} p{0.7cm} p{0.5cm} p{0.5cm} p{0.5cm} p{0.5cm} p{0.5cm} p{0.5cm}} 
\hline
$\rho$ & $2\%$& $4\%$& $6\%$& $8\%$& $10\%$& $12\%$& $14\%$& $16\%$\\
\hline
$T$& $\frac{1}{2}$& $\frac{1}{2}$& $\frac{2}{5}-\frac{1}{2}$&$\frac{3}{8}-\frac{3}{7}$& $\frac{3}{8}$&$\frac{3}{8}$&$\frac{3}{8}$&$\frac{1}{3}-\frac{3}{8}$\\
\hline
$\rho$ & $18\%$& $20\%$& $22\%$& $24\%$& $26\%$& $28\%$ & $30\%$& $32\%$\\
\hline\
$T$&$\frac{1}{3}-\frac{3}{8}$&$\frac{1}{3}$&$\frac{1}{4}-\frac{1}{3}$& $\frac{1}{4}-\frac{1}{3}$& $\frac{1}{5}-\frac{1}{4}$& $\frac{1}{5}-\frac{1}{4}$& $\frac{1}{5}$&  $\frac{1}{5}$\\
\hline
$\rho$& $34\%$& $36\%$& $38\%$& $40\%$& $42\%$& $44\%$& $46\%$\\
\hline
$T$& $\frac{1}{5}$&$\frac{1}{5}$& $\frac{1}{5}$&$\frac{1}{5}$&$\frac{1}{6}-\frac{1}{5}$&$\frac{1}{8}-\frac{1}{5}$&$\frac{1}{8}$\\
\hline
\end{tabular}
\label{tab}
\caption{''Frozen state-segregated state`` transition line. The limits between which the system may end in a frozen state or segregated state depending on the order of the dynamics are obtained by performing $100$ tests on which we look at the percentage of ''frozen states''. If for given $T$ and $\rho$, the percentage of frozen states (resp. segregated states) is very high ($>95\%$), we consider that the corresponding equilibrium configuration is a frozen one (resp. segregated).} 
\end{center}
\end{table}

\end{document}